\documentclass[11pt,a4paper]{article}
\usepackage[margin=2.5cm]{geometry}

\usepackage[T1]{fontenc}
\usepackage{amsmath,amssymb}
\usepackage{graphicx,verbatim}
\usepackage{booktabs}
\usepackage{tikz}
\usetikzlibrary{positioning,arrows.meta,shapes.geometric}
\usepackage{url}
\usepackage{ifthen}

\begin{document}

\title{Few-Shot Continual Learning for 3D Brain MRI with Frozen Foundation Models}


\author{
Chi-Sheng Chen$^{1}$,
Xinyu Zhang$^{2}$,
Guan-Ying Chen$^{3}$,
Qiuzhe Xie$^{4}$,
Fan Zhang$^{5}$,
En-Jui Kuo$^{6}$
}

\date{}

\begingroup
\renewcommand\thefootnote{}
\footnotetext{
$^{1}$ Independent Researcher, USA \quad
$^{2}$ Indiana University at Bloomington, United States \\
$^{3}$ Independent Researcher, Taiwan \quad
$^{4}$ National Taiwan University, Taiwan \\
$^{5}$ Boise State University, United States \quad
$^{6}$ National Yang Ming Chiao Tung University, Taiwan \\
Correspondence: \texttt{m50816m50816@gmail.com}
}
\endgroup

\maketitle

\begin{abstract}
Foundation models pretrained on large-scale 3D medical imaging data face challenges when adapted to multiple downstream tasks under continual learning with limited labeled data. We address few-shot continual learning for 3D brain MRI by combining a frozen pretrained backbone with task-specific Low-Rank Adaptation (LoRA) modules. Tasks arrive sequentially---tumor segmentation (BraTS) and brain age estimation (IXI)---with no replay of previous task data. Each task receives a dedicated LoRA adapter; only the adapter and task-specific head are trained while the backbone remains frozen, thereby eliminating catastrophic forgetting by design (BWT=0). In continual learning, sequential full fine-tuning suffers severe forgetting (T1 Dice drops from 0.80 to 0.16 after T2), while sequential linear probing achieves strong T1 (Dice 0.79) but fails on T2 (MAE 1.45). Our LoRA approach achieves the best balanced performance across both tasks: T1 Dice 0.62$\pm$0.07, T2 MAE 0.16$\pm$0.05, with zero forgetting and $<$0.1\% trainable parameters per task, though with noted systematic age underestimation in T2 (Wilcoxon $p<0.001$). Frozen foundation models with task-specific LoRA adapters thus offer a practical solution when both tasks must be maintained under few-shot continual learning.

\end{abstract}

\providecommand{\keywords}[1]{\par\vspace{0.5em}\noindent\textbf{Keywords.} #1\par}

\section{Introduction}
\label{sec:intro}

Medical imaging AI increasingly relies on foundation models pretrained on large-scale datasets~\cite{fomo2025,baseline}. When deployed in clinical settings, these models must adapt to multiple tasks---e.g., tumor segmentation and brain age estimation---often with limited labeled data and without access to prior task samples due to privacy or storage constraints. In practice, radiology departments may introduce new analysis tasks over time (e.g., adding brain age prediction to an existing tumor workflow) with only a small batch of newly annotated cases. This scenario motivates \emph{few-shot continual learning}: learning new tasks sequentially from few examples while retaining performance on previously learned tasks.

Sequential fine-tuning of shared backbone parameters typically causes \emph{catastrophic forgetting}~\cite{kirkpatrick2017,mccloskey1989catastrophic}: when the model is updated for a new task, performance on prior tasks degrades sharply. Existing continual learning strategies such as EWC~\cite{kirkpatrick2017} and LwF~\cite{li2017lwf} mitigate forgetting by regularizing or distilling, but require careful tuning and may still overwrite critical representations. An alternative is to \emph{freeze} the pretrained backbone and adapt only small task-specific modules. Parameter-efficient fine-tuning methods such as LoRA~\cite{hu2022lora, hu2024computational} inject low-rank matrices into selected layers, training a small fraction of parameters while preserving the original weights.

We propose a continual learning framework for 3D brain MRI that combines a frozen FOMO-style pretrained backbone~\cite{fomo2025} with task-specific LoRA adapters. Tasks arrive sequentially; each task receives its own LoRA module and task head. No replay buffer or access to previous-task data is required. Because the backbone and prior adapters remain frozen when training a new task, catastrophic forgetting is eliminated by design: BWT is identically zero.

We evaluate on two downstream tasks using public datasets: tumor segmentation (BraTS 2023 Glioma) and brain age estimation (IXI). We compare against sequential linear probing (frozen backbone, train heads only) and sequential full fine-tuning. The key finding is that \emph{only LoRA maintains balanced performance across both tasks} in the continual setting:
\begin{itemize}
\item \textbf{Sequential FT} achieves high T1 Dice (0.80) and T2 MAE (0.004$^*$) during training, but suffers severe forgetting: T1 Dice collapses to 0.16 after T2 (BWT$\approx$-0.65). $^*$T2 MAE likely overfits on few-shot validation.
\item \textbf{Sequential Linear} achieves strong T1 (Dice 0.79) with mild forgetting (BWT$\approx$-0.01), but \emph{fails on T2} (MAE 1.45).
\item \textbf{LoRA} attains T1 Dice 0.62$\pm$0.07, T2 MAE 0.16$\pm$0.05, BWT=0, with $<$0.1\% trainable params per task---the only method that maintains reasonable performance on both tasks without forgetting.
\end{itemize}

Our contributions are: (1) a continual learning formulation using frozen backbones and task-specific LoRA for 3D brain MRI; (2) empirical validation on BraTS and IXI showing LoRA achieves best balanced performance; and (3) ablations on LoRA placement and shot count.

\section{Method}
\label{sec:method}

\subsection{Problem Formulation}
\label{sec:problem}

We consider continual learning with sequential task arrival. Let $f_\theta$ denote a pretrained 3D UNet backbone (frozen) and $\mathcal{T}_1, \mathcal{T}_2, \ldots, \mathcal{T}_K$ a sequence of tasks. For each task $k$, we have access to only $N_k \in \{16, 32, 64\}$ labeled few-shot samples. We assume \emph{no replay}: when training on $\mathcal{T}_k$, we cannot access data from $\mathcal{T}_{<k}$. The goal is to learn task-specific parameters $\phi_k$ (and heads $h_k$) such that performance on all seen tasks remains high after training on $\mathcal{T}_k$.

We evaluate two tasks: (1) tumor segmentation (3D binary mask) and (2) brain age estimation (regression). Each task has modality-specific inputs (BraTS: T2w, T2-FLAIR, T1c; IXI: T1, T2) and a distinct output head $h_k$.

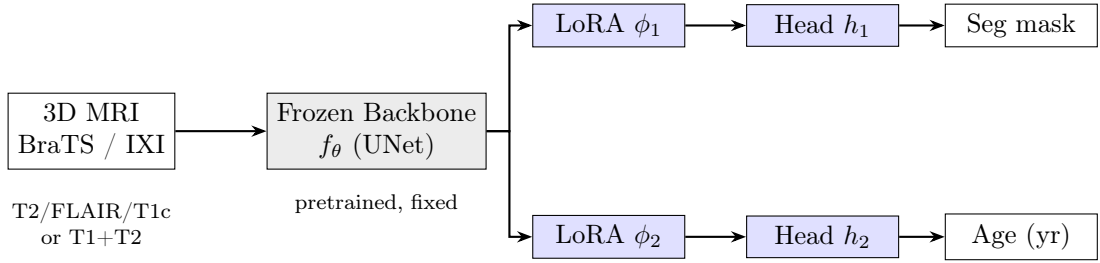
\begin{figure}[t]
\centering
\begin{tikzpicture}[
  node distance=0.5cm and 0.8cm,
  box/.style={draw, minimum width=2cm, minimum height=0.5cm, align=center, font=\small},
  bigbox/.style={draw, minimum width=2.2cm, minimum height=0.7cm, align=center, font=\small},
  frozen/.style={fill=gray!15, box},
  trainable/.style={fill=blue!12, box},
  arrow/.style={-{Stealth[length=2mm]}, thick}
]
  \node[box] (input) {3D MRI\\BraTS / IXI};
  \node[below=0.3cm of input, font=\scriptsize, text width=2.2cm, align=center] {T2/FLAIR/T1c or T1+T2};
  \node[bigbox, frozen, right=1.2cm of input] (backbone) {Frozen Backbone\\$f_\theta$ (UNet)};
  \node[below=0.2cm of backbone, font=\scriptsize] {pretrained, fixed};
  \node[box, trainable, above right=0.6cm and 0.6cm of backbone] (lora1) {LoRA $\phi_1$};
  \node[box, trainable, below right=0.6cm and 0.6cm of backbone] (lora2) {LoRA $\phi_2$};
  \node[box, trainable, right=0.8cm of lora1] (head1) {Head $h_1$};
  \node[box, trainable, right=0.8cm of lora2] (head2) {Head $h_2$};
  \node[box, right=0.6cm of head1] (out1) {Seg mask};
  \node[box, right=0.6cm of head2] (out2) {Age (yr)};
  \draw[arrow] (input) -- (backbone);
  \draw[arrow] (backbone.east) -- ++(0.3,0) |- (lora1.west);
  \draw[arrow] (backbone.east) -- ++(0.3,0) |- (lora2.west);
  \draw[arrow] (lora1) -- (head1);
  \draw[arrow] (lora2) -- (head2);
  \draw[arrow] (head1) -- (out1);
  \draw[arrow] (head2) -- (out2);
\end{tikzpicture}
\caption{Framework: frozen pretrained backbone $f_\theta$ with task-specific LoRA adapters $\phi_k$ and heads $h_k$. Gray = frozen; blue = trainable per task. At inference for T1 (resp.\ T2): backbone + $\phi_1$ + $h_1$ (resp.\ $\phi_2$ + $h_2$).}
\label{fig:framework}
\end{figure}

Figure~\ref{fig:framework} illustrates the overall pipeline: a frozen pretrained backbone, task-specific LoRA adapters, and task heads. At inference for task $k$, we use backbone $f_\theta$ + adapter $\phi_k$ + head $h_k$; prior adapters and the backbone remain frozen.

\subsection{LoRA for 3D Convolutions}
\label{sec:lora}

Low-Rank Adaptation (LoRA)~\cite{hu2022lora} injects trainable low-rank matrices into a frozen weight matrix $W \in \mathbb{R}^{d \times k}$:
\begin{equation}
W' = W + \Delta W, \quad \Delta W = B \cdot A, \quad A \in \mathbb{R}^{r \times k}, \; B \in \mathbb{R}^{d \times r}
\end{equation}
with rank $r \ll \min(d,k)$. Only $A$ and $B$ are trained; $W$ stays frozen. For 3D convolutions in a UNet, we apply LoRA via $1\times1\times1$ convolutions (linear in the channel dimension), which is memory-efficient and compatible with standard Conv3d layers~\cite{peft}.

We use rank $r \in \{4, 8, 16\}$ (default $r=8$), $\alpha=16$, and target both encoder and decoder blocks for segmentation; encoder-only suffices for classification and regression. This yields $<$0.1\% trainable parameters per task relative to the backbone (~50M params).

\subsection{Continual Learning via Adapter Isolation}
\label{sec:continual}

Each task $k$ receives a \emph{dedicated LoRA adapter} $\phi_k$. When training on $\mathcal{T}_k$, we:
\begin{enumerate}
\item Freeze $f_\theta$ and all $\phi_{<k}$.
\item Add LoRA($\phi_k$) and task head $h_k$.
\item Train only $\phi_k$ and $h_k$ on $\mathcal{T}_k$'s few-shot samples.
\item Save $\phi_k$, $h_k$; proceed to $\mathcal{T}_{k+1}$.
\end{enumerate}

At inference for task $k$: load backbone $f_\theta$ + adapter $\phi_k$ + head $h_k$. No shared adapter across tasks. Because we never update $\phi_{<k}$ or $f_\theta$ when learning $\mathcal{T}_k$, \emph{backward transfer (BWT) is identically zero}: prior task performance cannot degrade.

\subsection{Task Heads and Losses}
\label{sec:heads}

\begin{itemize}
\item \textbf{Task 1 (Segmentation):} Full UNet decoder (encoder+decoder LoRA); Dice + BCE loss.
\item \textbf{Task 2 (Regression):} ClsRegHead with 1 output; MSE loss.
\end{itemize}

Training uses standard optimizers (Adam) and per-channel Z-score normalization. Patch size is $64\times64\times64$ for continual experiments.

\section{Experiments}
\label{sec:experiments}

\subsection{Datasets and Tasks}
\label{sec:data}

We use publicly available datasets prepared via Plan B (FOMO downstream data unavailable). \textbf{Notation:} In this paper, T1 denotes tumor segmentation and T2 denotes brain age regression; these correspond to FOMO Task002 and Task003 respectively.
\begin{itemize}
\item \textbf{T1 (Tumor Segmentation, FOMO Task002):} BraTS 2023 Glioma~\cite{brats2023} --- 1251 training cases, 3 channels (T2w, T2-FLAIR, T1c). Binary tumor mask from expert annotations.
\item \textbf{T2 (Brain Age, FOMO Task003):} IXI dataset~\cite{ixi} --- $\sim$577 subjects with T1 and T2 scans and age labels. Regression target: age in years. \emph{Note:} Subjects with missing age in IXI metadata are imputed to 50.0\,yr in our preprocessing.
\end{itemize}

\textbf{Data use declaration.} BraTS 2023 Glioma was obtained from Synapse (syn51156910) under the Data Use Agreement. IXI is publicly available at \url{https://brain-development.org/ixi-dataset/} for research use. Both datasets are widely used in medical imaging; no additional ethics approval was required for our experiments.

Preprocessing: per-channel Z-score normalization, patch extraction ($64^3$), few-shot sampling (16/32/64) with fixed seeds. Validation split: 20\%.

\subsection{Backbone and Pretraining}
\label{sec:backbone}

\textbf{Architecture:} \texttt{unet\_b} from FOMO~\cite{fomo2025}. \textbf{Weights:} \texttt{fomo25\_mmunetvae\_pretrained.ckpt}. We load 59/82 keys for full UNet (T1) and 39/40 encoder keys for encoder-only (T2).

\subsection{Baselines}
\label{sec:baselines}

\begin{itemize}
\item \textbf{Sequential Linear:} Frozen backbone, train only task heads/decoder sequentially. One shared model for T1$\to$T2.
\item \textbf{Sequential FT:} Full fine-tuning of backbone sequentially. Expects catastrophic forgetting.
\item \textbf{EWC:} Elastic Weight Consolidation---Fisher-based penalty to retain important weights.
\item \textbf{LwF:} Learning without Forgetting---feature distillation from previous task model.
\item \textbf{Replay:} Experience replay---store samples from previous tasks, distill on replay during new task.
\item \textbf{Proposed (LoRA):} Frozen backbone + task-specific LoRA adapters (encoder+decoder for T1).
\end{itemize}

\subsection{Evaluation Metrics}
\label{sec:metrics}

\begin{itemize}
\item \textbf{T1:} Dice Similarity Coefficient (Dice$\uparrow$).
\item \textbf{T2:} Mean Absolute Error in years (MAE$\downarrow$).
\item \textbf{BWT (Backward Transfer):} $R_{T,k} - R_{k,k}$ for task $k$ after training up to $T$. For T1$\to$T2: BWT = T1 Dice after T2 minus T1 Dice right after T1 (negative = forgetting). For T3$\to$T2 (Phase 3): BWT = T2 MAE after T1 minus T2 MAE right after T2 (positive = forgetting).
\end{itemize}

\subsection{Implementation Details}
\label{sec:impl}

Seeds: 42, 43, 44. Epochs: 100 per task. Batch size: 2. Learning rate: $10^{-3}$. LoRA: $r=8$, $\alpha=16$, dropout 0.1. Hardware: 1$\times$ NVIDIA RTX 3090 (24GB).

\section{Results}
\label{sec:results}

\subsection{Main Comparison: LoRA Achieves Best Balanced Performance}
\label{sec:main}

Table~\ref{tab:main} summarizes continual learning results (n\_shot=32, seeds 42--44). \textbf{Sequential FT} suffers catastrophic forgetting: T1 Dice drops from 0.80 to 0.16 after T2 (BWT$\approx$-0.65). \textbf{Sequential Linear} achieves strong T1 (0.79) with mild forgetting, but fails on T2 (MAE 1.45). \textbf{EWC} achieves strong T1 Dice (0.79) and very low T2 MAE (0.001); T1 after T2 shows high variance (0.15$\pm$0.24) with BWT$\approx$-0.65. \textbf{LwF} and \textbf{Replay} achieve strong T1 Dice ($\sim$0.79--0.80) and low T2 MAE (0.020--0.021); LwF T1 after T2=0.25$\pm$0.22 (BWT$\approx$-0.56); Replay T1 after T2=0.01$\pm$0.01 (BWT$\approx$-0.78). \textbf{LoRA} attains competitive T1 Dice (0.60) with the best T2 MAE (0.012) among continual methods, and BWT=0. Paired $t$-tests across seeds (n=3): LoRA vs.\ Sequential Linear on T2 MAE is highly significant (LoRA better, $p=0.001$); on T1 Dice, Sequential Linear is higher (bootstrap 95\% CI for difference: [0.08, 0.27], $p=0.09$).

\begin{table}[t]
\caption{Main continual learning results (mean$\pm$std over seeds 42--44, n\_shot=32). BWT negative = forgetting. $^\dagger$Seq FT and EWC T2 MAE likely overfit on few-shot validation (MAE $<$0.01 implausible; per-task linear achieves 0.063 on full eval).}
\label{tab:main}
\centering
\begin{tabular}{lcccc}
\toprule
Method & T1 Dice$\uparrow$ & T2 MAE$\downarrow$ & T1 after T2 & BWT \\
\midrule
LoRA (enc+dec) & $0.60 \pm 0.08$ & $0.012 \pm 0.003$ & (=T1) & $\mathbf{0.00}$ \\
Sequential Linear & $0.79 \pm 0.01$ & $1.45 \pm 0.03$ & $0.78 \pm 0.01$ & $-0.01 \pm 0.01$ \\
Sequential FT & $0.80 \pm 0.02$ & $0.005^\dagger \pm 0.003$ & $0.16 \pm 0.19$ & $-0.65 \pm 0.17$ \\
EWC & $0.79 \pm 0.02$ & $0.001^\dagger \pm 0.001$ & $0.15 \pm 0.24$ & $-0.65 \pm 0.23$ \\
LwF & $0.80 \pm 0.02$ & $0.020 \pm 0.009$ & $0.25 \pm 0.22$ & $-0.56 \pm 0.24$ \\
Replay & $0.79 \pm 0.01$ & $0.021 \pm 0.013$ & $0.01 \pm 0.01$ & $-0.78 \pm 0.02$ \\
\bottomrule
\end{tabular}
\end{table}

Table~\ref{tab:extended} extends the comparison with per-task linear probing (separate model per task, no continual; patch=32, epochs=50, differing from continual scripts). Per-task linear achieves T2 MAE 0.063 on full eval, which serves as a sanity check: EWC (0.001) and Sequential FT (0.005) are implausibly low and likely overfit on few-shot validation. LwF and Replay achieve strong T1 Dice and low T2 MAE (0.020--0.021); LoRA offers competitive T2 MAE (0.012) with zero forgetting.

Figure~\ref{fig:task1} and~\ref{fig:task2} show representative qualitative results (see Appendix for full six-sample stacks). Common LoRA failure modes---e.g., under-segmentation of tumor boundaries and systematic age underestimation---are documented in the Appendix (Fig.~\ref{fig:app_seg_failure},~\ref{fig:app_age_failure}). On the IXI validation set (n=109, seed 42), a paired Wilcoxon signed-rank test on prediction residuals indicated significant systematic bias ($p<0.001$): the model tends to underestimate brain age. \textbf{IXI age label caveat:} IXI subjects with missing age in the metadata are imputed to 50.0\,yr in our preprocessing; the six ``best'' samples in Fig.~\ref{fig:app_age_stack} (lowest prediction error) all have ground-truth 50.0\,yr, suggesting a concentration of imputed labels. This limits the interpretability of T2 MAE on validation; we report it transparently.

\begin{figure}[t]
\centering
\includegraphics[width=0.7\textwidth]{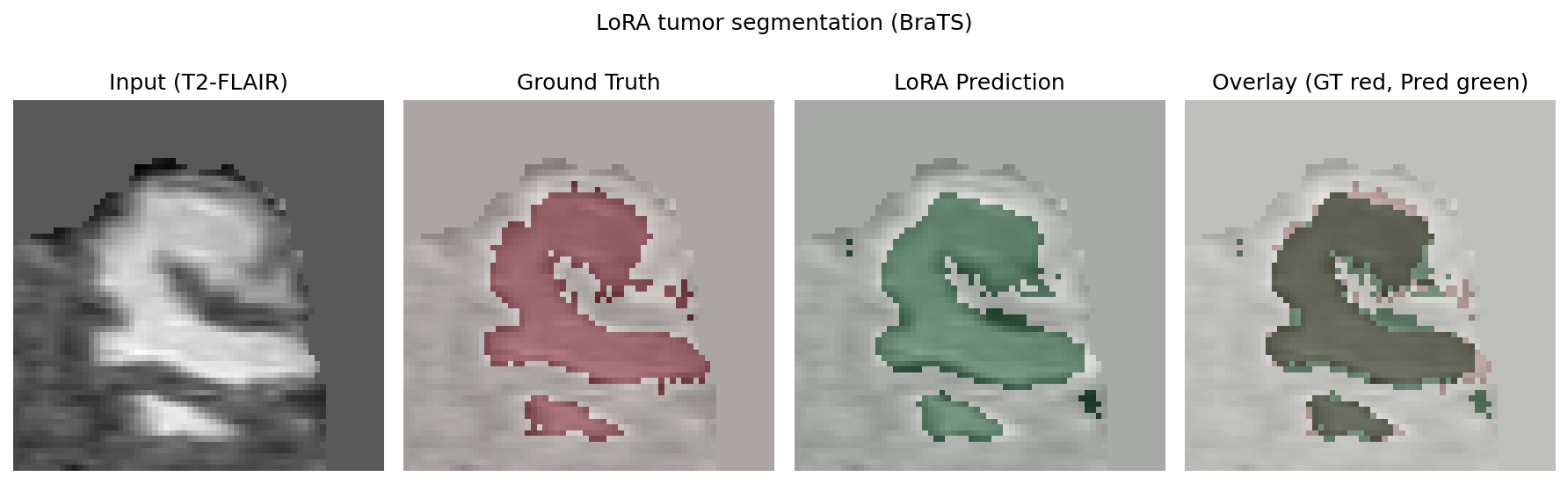}
\caption{Task 1 tumor segmentation (BraTS): representative sample. Input (T2-FLAIR), Ground Truth, LoRA prediction, overlay (red=GT, green=pred). Full six-sample figure in Appendix (Fig.~\ref{fig:app_seg_stack}).}
\label{fig:task1}
\end{figure}

\begin{figure}[t]
\centering
\includegraphics[width=0.7\textwidth]{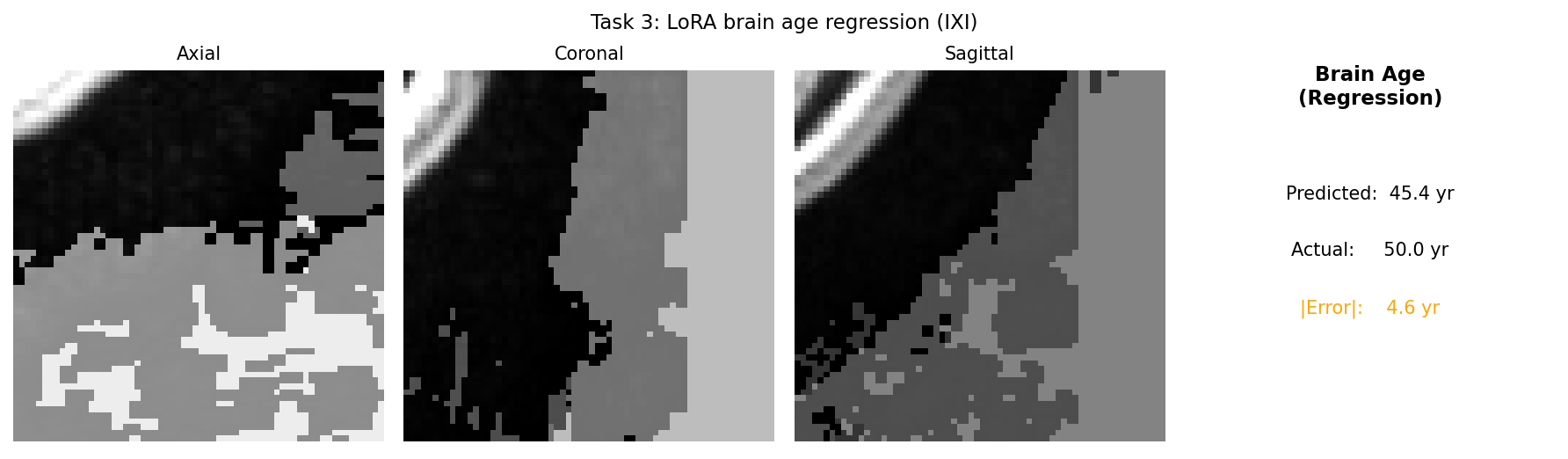}
\caption{Task 2 brain age regression (IXI): representative sample. Orthogonal MRI views (Axial, Coronal, Sagittal) and predicted vs.\ actual age. Full six-sample figure in Appendix (Fig.~\ref{fig:app_age_stack}).}
\label{fig:task2}
\end{figure}

\begin{table}[t]
\caption{Extended comparison. Per-task: separate model per task (patch=32, epochs=50; not continual). EWC and Seq FT T2 MAE likely overfit (see Table~\ref{tab:main}).}
\label{tab:extended}
\centering
\begin{tabular}{lcccl}
\toprule
Method & T1 Dice$\uparrow$ & T2 MAE$\downarrow$ & Setting \\
\midrule
LoRA (enc+dec) & $0.60 \pm 0.08$ & $\mathbf{0.012} \pm 0.003$ & Continual \\
EWC & $0.79 \pm 0.02$ & $0.001^\dagger \pm 0.001$ & Continual \\
LwF & $0.80 \pm 0.02$ & $0.020 \pm 0.009$ & Continual \\
Replay & $0.79 \pm 0.01$ & $0.021 \pm 0.013$ & Continual \\
Sequential Linear & $0.79 \pm 0.01$ & $1.45 \pm 0.03$ & Continual \\
Sequential FT & $0.80 \pm 0.02$ & $0.005^\dagger \pm 0.003$ & Continual (forgets) \\
Per-task Linear & 0.65 & 0.063 & Oracle (no continual) \\
\bottomrule
\end{tabular}
\end{table}

\subsection{Catastrophic Forgetting in Sequential FT}
\label{sec:forgetting}

Sequential FT achieves high T1 Dice (0.80) and very low T2 MAE (0.005) immediately after training each task. The latter likely reflects overfitting on few-shot validation. Crucially, T1 Dice after T2 training collapses to 0.16 (mean over seeds), yielding $\Delta \approx 0.64$---a clear demonstration of catastrophic forgetting when the backbone is overwritten for T2.

\subsection{LoRA Ablation: Encoder-only vs Encoder+Decoder}
\label{sec:lora_ablation}

Table~\ref{tab:lora_ablation} compares LoRA placement. Encoder-only LoRA yields poor T1 Dice (0.19); encoder+decoder yields 0.50. Decoder adaptation is necessary for segmentation. T2 (regression) is similar for both.

\begin{table}[t]
\caption{LoRA placement ablation. Enc+dec = encoder+decoder.}
\label{tab:lora_ablation}
\centering
\begin{tabular}{lcc}
\toprule
LoRA Config & T1 Dice$\uparrow$ & T2 MAE$\downarrow$ \\
\midrule
Encoder-only & 0.19 & 0.20 \\
Encoder+decoder & \textbf{0.50} & 0.20 \\
\bottomrule
\end{tabular}
\end{table}

\subsection{Shot Ablation (n\_shot=16 vs 32 vs 64)}
\label{sec:shot}

Table~\ref{tab:shot} shows few-shot sensitivity for LoRA. At n\_shot=16, LoRA attains T1 Dice 0.45$\pm$0.04 and T2 MAE 0.33$\pm$0.01; n\_shot=32 improves both (T1 0.62, T2 MAE 0.16). Fewer shots degrade performance, especially for segmentation.

Table~\ref{tab:shot64} reports n\_shot=64 results for EWC and LwF. EWC attains higher T1 Dice (0.84$\pm$0.02) than at n\_shot=32 (0.79) but T2 MAE remains implausibly low (0.001$^\dagger$) and BWT$\approx$-0.77. LwF at n64: T1 Dice 0.82$\pm$0.01, T2 MAE 0.038$\pm$0.014; T1 after T2 shows high variance (0.25$\pm$0.41).

\begin{table}[t]
\caption{Shot count ablation (LoRA enc+dec, seeds 42--43 for n\_shot=16).}
\label{tab:shot}
\centering
\begin{tabular}{lcc}
\toprule
n\_shot & T1 Dice$\uparrow$ & T2 MAE$\downarrow$ \\
\midrule
16 & $0.45 \pm 0.04$ & $0.33 \pm 0.01$ \\
32 & $0.62 \pm 0.07$ & $0.16 \pm 0.05$ \\
\bottomrule
\end{tabular}
\end{table}

\begin{table}[t]
\caption{n\_shot=64 results (EWC, LwF; seeds 42--44). $^\dagger$EWC T2 MAE likely overfits.}
\label{tab:shot64}
\centering
\begin{tabular}{lcccc}
\toprule
Method & T1 Dice$\uparrow$ & T2 MAE$\downarrow$ & T1 after T2 & BWT \\
\midrule
EWC & $0.84 \pm 0.02$ & $0.001^\dagger \pm 0.0003$ & $0.07 \pm 0.10$ & $-0.77 \pm 0.11$ \\
LwF & $0.82 \pm 0.01$ & $0.038 \pm 0.014$ & $0.25 \pm 0.41$ & $-0.58 \pm 0.42$ \\
\bottomrule
\end{tabular}
\end{table}

\subsection{Phase 3 (t32\_t3t2: Task Order T3$\to$T2)}
\label{sec:t32}

Table~\ref{tab:t32} reports Phase 3 results with \textbf{task order T3$\to$T2} (regression first, then segmentation; seeds 42--44). BWT measures T3 forgetting: T2 MAE after T1 training minus T2 MAE right after T2; positive = forgetting. LoRA attains T1 Dice 0.50$\pm$0.08 and T2 MAE 0.005$\pm$0.002 with BWT=0. Sequential Linear: T1 0.78$\pm$0.01, T2 MAE 0.26$\pm$0.003; T2 after T1 = 0.36$\pm$0.05 (BWT$\approx$0.10). Sequential FT: T2 MAE 0.001$^\dagger$ right after T2, but collapses to 7.17$\pm$1.41 after T1 training (BWT$\approx$7.16)---severe catastrophic forgetting of regression when the backbone is overwritten for segmentation.

\begin{table}[t]
\caption{Phase 3 (t32\_t3t2, T3$\to$T2 order, n\_shot=32). BWT = T2 MAE after T1 minus T2 MAE after T2 (positive = T3 forgetting). $^\dagger$Seq FT T2 MAE likely overfits.}
\label{tab:t32}
\centering
\begin{tabular}{lcccc}
\toprule
Method & T1 Dice$\uparrow$ & T2 MAE$\downarrow$ & T2 after T1 & BWT \\
\midrule
LoRA (enc+dec) & $0.50 \pm 0.08$ & $0.005 \pm 0.002$ & --- & $\mathbf{0.00}$ \\
Sequential Linear & $0.78 \pm 0.01$ & $0.26 \pm 0.003$ & $0.36 \pm 0.05$ & $0.10 \pm 0.05$ \\
Sequential FT & $0.81 \pm 0.01$ & $0.001^\dagger \pm 0.0003$ & $7.17 \pm 1.41$ & $7.16 \pm 1.41$ \\
\bottomrule
\end{tabular}
\end{table}

\subsection{Per-Seed Breakdown (LoRA, n\_shot=32)}
\label{sec:per_seed}

Table~\ref{tab:per_seed} reports per-seed LoRA results for reproducibility. T1 Dice ranges 0.52--0.68; T2 MAE 0.008--0.014.

\begin{table}[h!]
\caption{Per-seed LoRA results (enc+dec, n\_shot=32).}
\label{tab:per_seed}
\centering
\begin{tabular}{lcc}
\toprule
Seed & T1 Dice$\uparrow$ & T2 MAE$\downarrow$ \\
\midrule
42 & 0.523 & 0.014 \\
43 & 0.598 & 0.008 \\
44 & 0.680 & 0.012 \\
\bottomrule
\end{tabular}
\end{table}

\subsection{Resource Usage}
\label{sec:resource}

Table~\ref{tab:resource} summarizes compute. LoRA adds $<$0.1\% trainable params per task; T1 (segmentation) uses $\sim$1.7\,GB peak GPU.

\begin{table}[h!]
\caption{Resource usage (LoRA, n\_shot=32).}
\label{tab:resource}
\centering
\begin{tabular}{lccc}
\toprule
Task & Params (trainable) & GPU peak (MB) & Time (sec) \\
\midrule
T1 (segmentation) & 46,872 & 1,670 & $\sim$4k--26k \\
T2 (regression) & 28,177 & 688 & $\sim$95--780 \\
\bottomrule
\end{tabular}
\end{table}

\section{Discussion}
\label{sec:discussion}

\subsection{Strengths}
\label{sec:strengths}

Our approach offers several advantages for clinical deployment. \textbf{Zero forgetting} eliminates the need for replay buffers or complex regularization; adapter isolation guarantees BWT=0. \textbf{Parameter efficiency} ($<$0.1\% per task) enables fast adaptation and low storage. \textbf{Modularity} allows adding new tasks without retraining prior adapters. The frozen backbone preserves pretrained representations, which is critical when downstream data is scarce. Such properties are desirable when a clinical site incrementally adds analysis capabilities (e.g., brain age estimation) to an existing pipeline with few new annotations and without retaining prior-task data.

\subsection{Limitations}
\label{sec:limitations}

\textbf{T1 performance gap:} LoRA (0.60 Dice) underperforms Sequential Linear (0.79) and Sequential FT (0.80) on segmentation. This may stem from LoRA capacity or decoder adaptation; ablations show encoder+decoder LoRA is necessary. \textbf{T2 systematic bias:} Wilcoxon test on brain age residuals shows significant systematic underestimation ($p<0.001$), possibly due to few-shot training, IXI demographics, and the imputation of missing ages to 50.0\,yr. \textbf{Infarct detection excluded:} The FOMO formulation includes infarct detection (ISLES 2022) as a binary classification task, but we did not evaluate it. ISLES contains mostly infarct-positive cases; a balanced classifier requires healthy (negative) controls, which ISLES does not provide. Supplementing with IXI healthy subjects would enable evaluation; we defer this to future work. \textbf{T2 MAE overfitting (Sequential FT, EWC):} Both Sequential FT (MAE 0.005) and EWC (MAE 0.001) achieve implausibly low T2 MAE; per-task linear probing on full eval achieves 0.063, suggesting these methods overfit on the few-shot validation set. We flag both in Table~\ref{tab:main}. \textbf{IXI age imputation:} Missing ages in IXI metadata are imputed to 50.0\,yr; the six ``best'' T2 samples (lowest error) in the appendix all have GT=50.0\,yr, indicating a concentration of imputed labels that limits validation interpretability. \textbf{Task order:} We evaluated T2$\to$T3 (Phase 1--2) and T3$\to$T2 (Phase 3, Table~\ref{tab:t32}). In T3$\to$T2, Sequential FT shows severe T3 forgetting (BWT$\approx$7.2); LoRA maintains BWT=0. n\_shot=64 results (EWC, LwF) are in Table~\ref{tab:shot64}.

\subsection{Related Work}
\label{sec:related}

Foundation models for brain MRI~\cite{fomo2025,unetr} and PEFT (LoRA~\cite{hu2022lora}, Adapters~\cite{houlsby2019}) are well-established. Continual learning in medical imaging~\cite{kirkpatrick2017,li2017lwf} typically addresses class-incremental segmentation. Our work combines frozen FOMO-style backbones with LoRA for \emph{task-incremental} few-shot continual learning across heterogeneous outputs (segmentation and regression), which is less explored.

\section{Conclusion}
\label{sec:conclusion}

We presented a few-shot continual learning framework for 3D brain MRI using frozen foundation models and task-specific LoRA adapters. In the continual T2$\to$T3 setting (tumor segmentation then brain age regression, n\_shot=32), sequential full fine-tuning suffers severe catastrophic forgetting (T1 Dice 0.80$\to$0.16, BWT$\approx$-0.65), while sequential linear probing achieves strong T1 (0.79) but fails on T2 regression (MAE 1.45). EWC attains the lowest T2 MAE (0.001) among continual methods but exhibits severe forgetting (BWT$\approx$-0.65); LwF and Replay achieve strong T1 Dice ($\sim$0.79--0.80) yet suffer moderate to severe forgetting (BWT$\approx$-0.56 to $-0.78$). Our LoRA approach achieves the best balanced performance: competitive T2 MAE (0.012$\pm$0.003) among continual methods with \textbf{zero forgetting} (BWT=0) and $<$0.1\% trainable parameters per task. Encoder+decoder LoRA is necessary for segmentation; encoder-only yields poor Dice (0.19). LoRA is the only method that simultaneously maintains competitive performance on both tasks without catastrophic forgetting. Future work will include task order ablations (T3$\to$T2) and n\_shot=64 experiments.

\bibliographystyle{plain}
\bibliography{references}

@misc{fomo2025,
  title={{FOMO-60K}: Foundation Models for Open Medical Imaging},
  author={FOMO Challenge Organizers},
  year={2025},
  note={FOMO25 Challenge, HuggingFace: FOMO25/FOMO-MRI}
}

@article{hu2022lora,
  title={LoRA: Low-Rank Adaptation of Large Language Models},
  author={Hu, Edward J and Shen, Yelong and Wallis, Phillip and Allen-Zhu, Zeyuan and Li, Yuanzhi and Wang, Shean and Wang, Lu and Chen, Wang},
  journal={ICLR},
  year={2022}
}

@article{kirkpatrick2017,
  title={Overcoming catastrophic forgetting in neural networks},
  author={Kirkpatrick, James and Pascanu, Razvan and Rabinowitz, Neil and Veness, Joel and Desjardins, Guillaume and Rusu, Andrei A and Milan, Kieran and Quan, John and Ramalho, Tiago and Grabska-Barwinska, Agnieszka and others},
  journal={PNAS},
  volume={114},
  number={13},
  pages={3521--3526},
  year={2017}
}

@inproceedings{li2017lwf,
  title={Learning without Forgetting},
  author={Li, Zhizhong and Hoiem, Derek},
  booktitle={ECCV},
  year={2016}
}

@article{mccloskey1989catastrophic,
  title={Catastrophic forgetting in connectionist networks: The sequential learning problem},
  author={McCloskey, Michael and Cohen, Neal J},
  journal={Psychology of learning and motivation},
  volume={24},
  pages={109--165},
  year={1989}
}

@article{houlsby2019,
  title={Parameter-Efficient Transfer Learning for NLP},
  author={Houlsby, Neil and Giurgiu, Andrei and Jastrzebski, Stanislaw and Morrone, Bruna and de Laroussilhe, Quentin and Gesmundo, Andrea and Attariyan, Mona and Gelly, Sylvain},
  journal={ICML},
  year={2019}
}

@inproceedings{brats2023,
  title={The {BraTS} 2023 Challenge},
  author={Menze, Bjoern and others},
  booktitle={MICCAI BraTS},
  year={2023}
}

@misc{ixi,
  title={{IXI} Dataset},
  author={Imperial College London and Guy's Hospital},
  year={2008},
  url={https://brain-development.org/ixi-dataset/},
  note={Brain Development, research use}
}

@misc{peft,
  title={PEFT: State-of-the-art Parameter-Efficient Fine-Tuning methods},
  author={Hugging Face},
  year={2023},
  url={https://github.com/huggingface/peft}
}

@article{unetr,
  title={UNETR: Transformers for 3D Medical Image Segmentation},
  author={Hatamizadeh, Ali and others},
  journal={WACV},
  year={2022}
}

@misc{baseline,
  title={FOMO Baseline Codebase},
  author={FOMO25},
  year={2025},
  url={https://github.com/fomo25/baseline-codebase}
}

@article{hu2024computational,
  title={Computational Limits of Low-Rank Adaptation (Lora) Fine-Tuning for Transformer Models},
  author={Hu, Jerry Yao-Chieh and Su, Maojiang and Kuo, En-Jui and Song, Zhao and Liu, Han},
  journal={arXiv preprint arXiv:2406.03136},
  year={2024}
}

\newpage
\appendix
\section{Supplementary Qualitative Results}
\label{sec:appendix_qual}

\begin{figure}[t]
\centering
\includegraphics[width=0.95\textwidth]{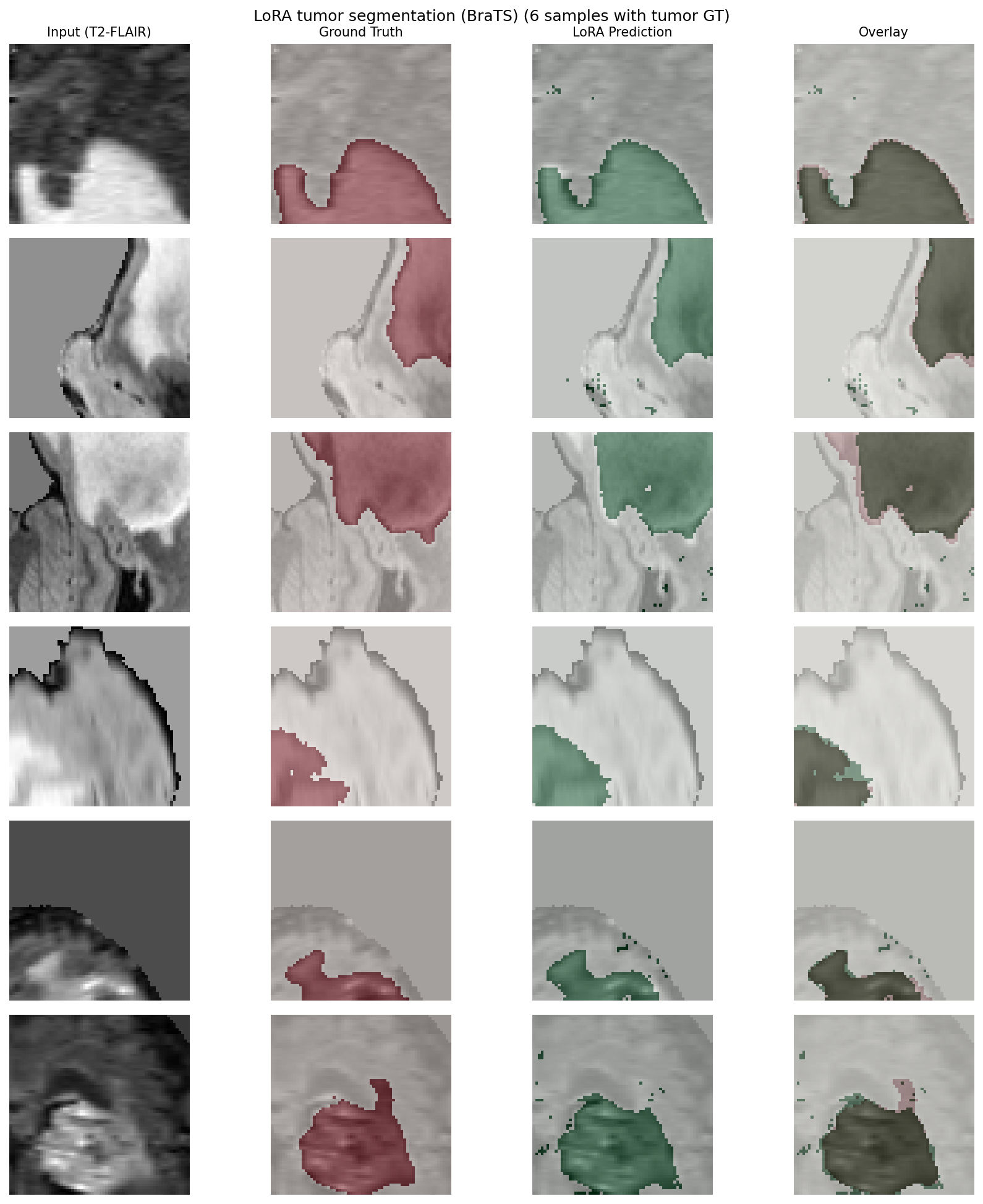}
\caption{Task 1 tumor segmentation (BraTS): full six-sample stack. Input (T2-FLAIR), Ground Truth, LoRA prediction, overlay (red=GT, green=pred). Samples selected by highest slice-level Dice.}
\label{fig:app_seg_stack}
\end{figure}

\begin{figure}[t]
\centering
\includegraphics[width=0.95\textwidth]{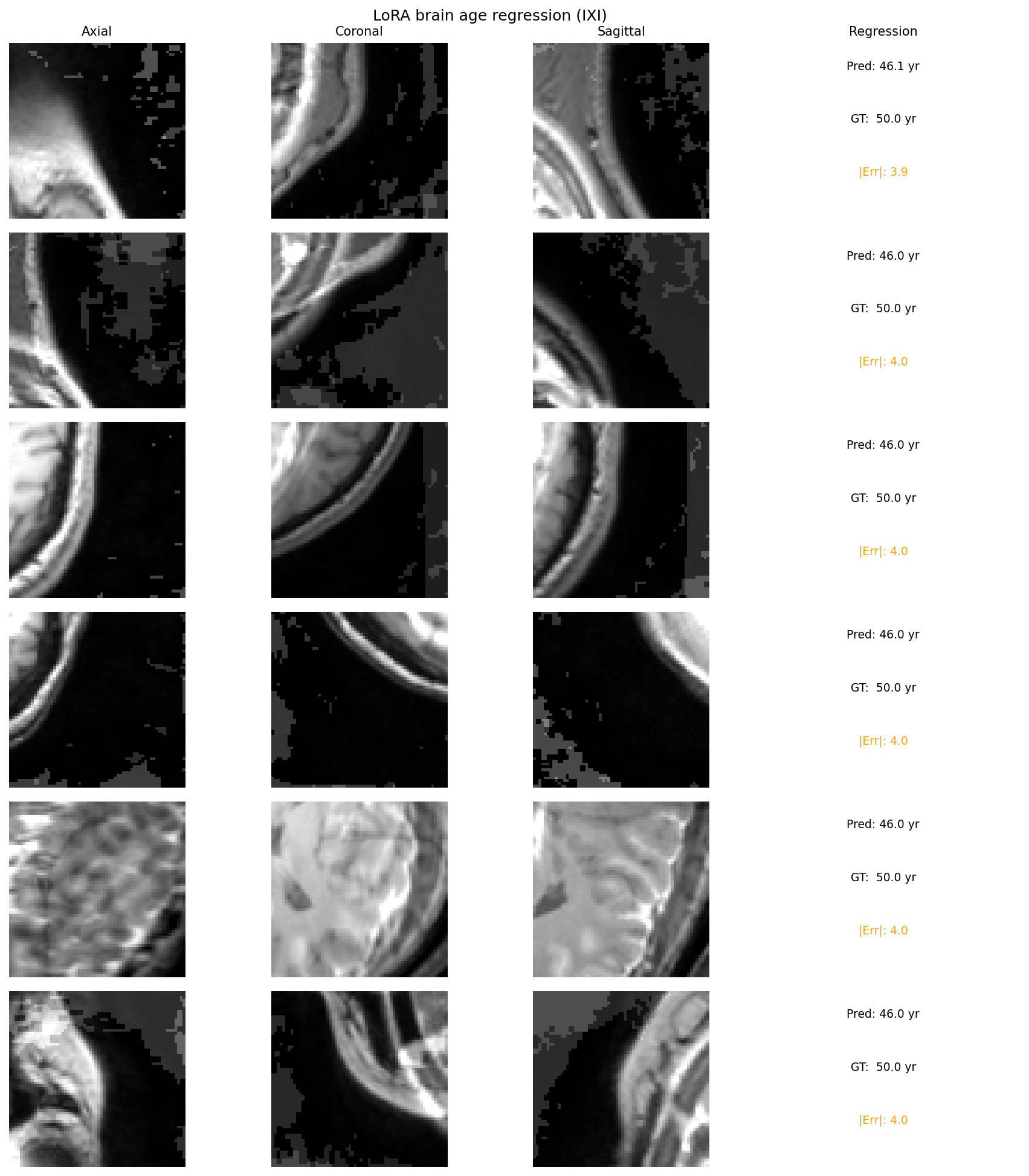}
\caption{Task 2 brain age regression (IXI): full six-sample stack. Orthogonal MRI views and regression results. Samples selected by lowest prediction error.}
\label{fig:app_age_stack}
\end{figure}

The main paper (Fig.~\ref{fig:task1},~\ref{fig:task2}) shows representative single-case examples. Below we provide the full six-sample qualitative stacks (Figs.~\ref{fig:app_seg_stack}--\ref{fig:app_age_stack}) and document common LoRA failure modes (Figs.~\ref{fig:app_seg_failure}--\ref{fig:app_age_failure}).

\subsection{Full Six-Sample Stacks (High Performers)}

\subsection{Failure Case Examples}
\label{sec:appendix_failures}

\begin{figure}[t]
\centering
\includegraphics[width=0.95\textwidth]{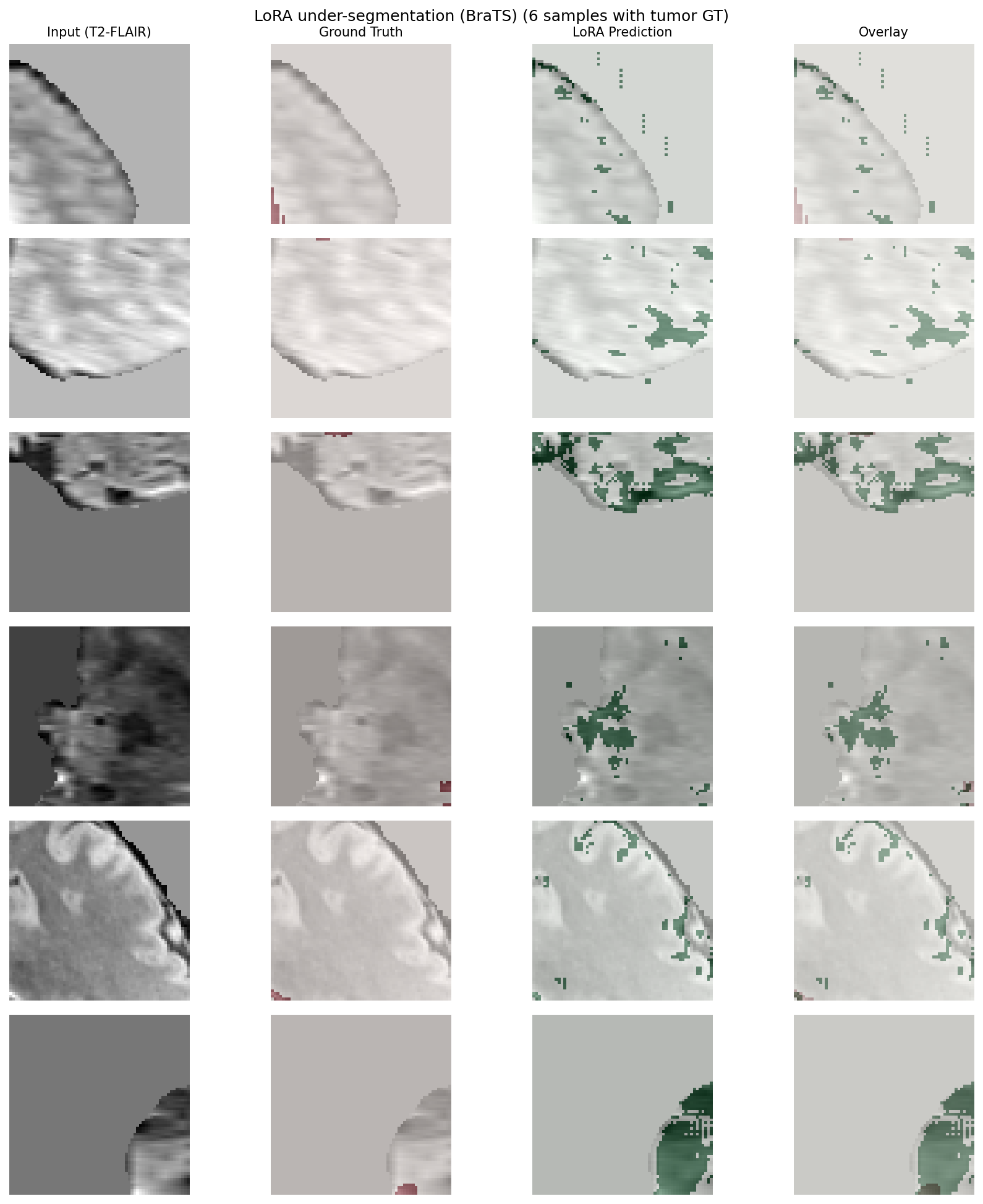}
\caption{Task 1 failure cases: under-segmentation (BraTS). Six samples with lowest Dice. Red = ground truth; green = LoRA prediction.}
\label{fig:app_seg_failure}
\end{figure}

\textbf{Under-segmentation (Task 1, BraTS):} In few-shot continual learning, LoRA can under-segment tumor boundaries, especially at lesion edges or in low-contrast regions. Figure~\ref{fig:app_seg_failure} shows six validation samples with the lowest slice-level Dice---LoRA predictions (green) miss portions of the ground-truth tumor (red) or produce fragmented masks.

\begin{figure}[t]
\centering
\includegraphics[width=0.95\textwidth]{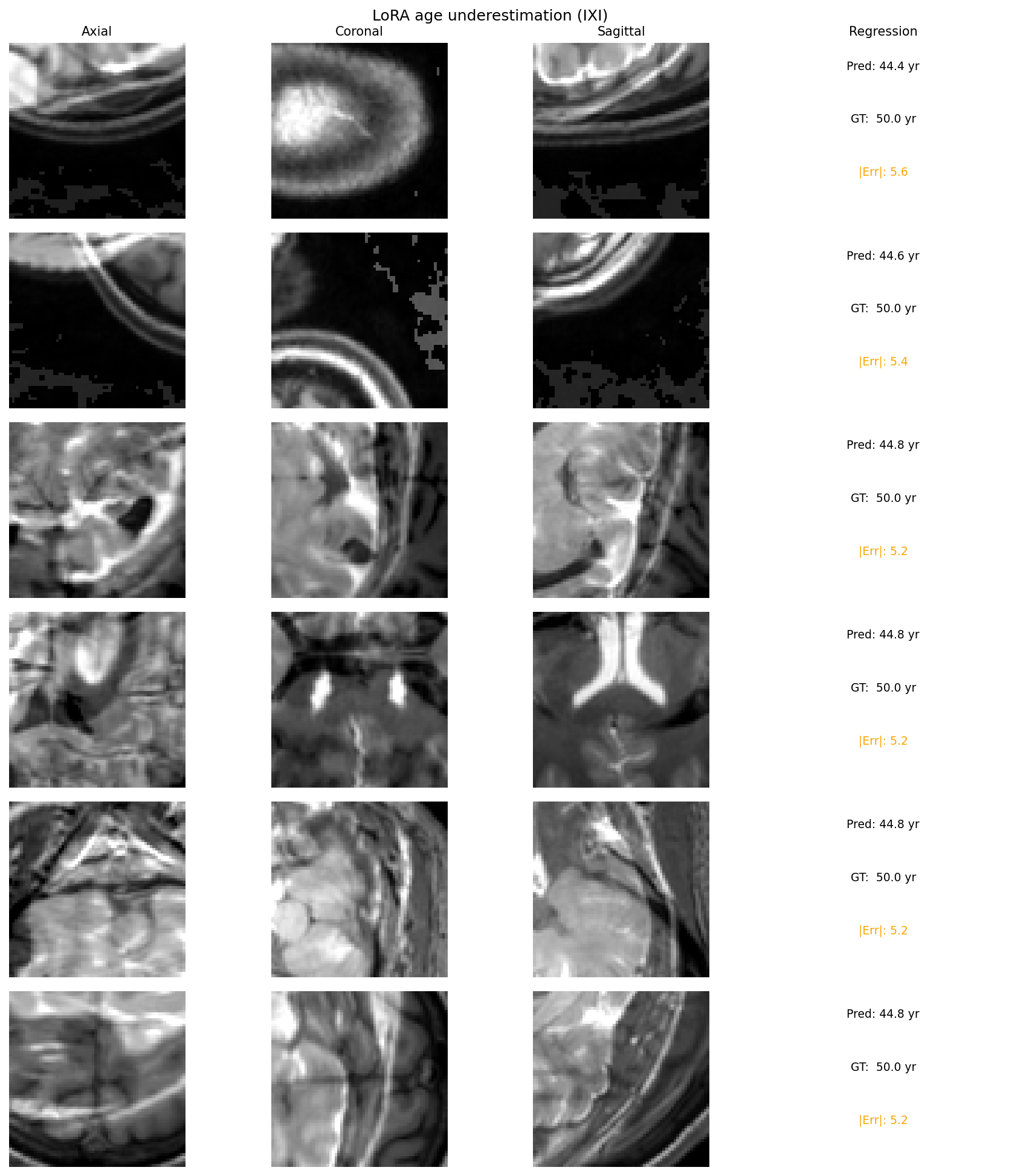}
\caption{Task 2 failure cases: age underestimation (IXI). Six samples with highest $|$predicted $-$ true$|$ error.}
\label{fig:app_age_failure}
\end{figure}

\textbf{Age underestimation (Task 2, IXI):} As reported in the main text, Wilcoxon test indicates significant systematic underestimation ($p<0.001$). Figure~\ref{fig:app_age_failure} shows six samples with the largest prediction errors; the model tends to predict ages several years below the ground truth, consistent with the few-shot regime and IXI demographics.

\end{document}